%% file: skeleton.tex
\documentclass{PoS}

\usepackage{graphicx}
\usepackage{dcolumn}% Align table columns on decimal point
\usepackage{float}
\usepackage{amsmath}
\usepackage{amsfonts}
\usepackage{multirow}
\usepackage{color}
\usepackage{wasysym}
\usepackage{soul}
\usepackage{slashed}

\def\GeV{\ifmmode {\mathrm{\ Ge\kern -0.1em V}}\else
                   \textrm{Ge\kern -0.1em V}\fi}%
\def\TeV{\ifmmode {\mathrm{\ Te\kern -0.1em V}}\else
                   \textrm{Te\kern -0.1em V}\fi}%

\newcommand{\beq}{\begin{equation}}
\newcommand{\eeq}{\end{equation}}
\newcommand{\beqn}{\begin{eqnarray}}
\newcommand{\eeqn}{\end{eqnarray}}

\newcommand{\nn}{\nonumber}

\newcommand{\ora}{\overrightarrow}

\newcommand{\leftb}{\left(}
\newcommand{\rightb}{\right)}

\def\beq{\begin{equation}}
\def\eeq{\end{equation}}
\def\beqn{\begin{eqnarray}}
\def\eeqn{\end{eqnarray}}
\def\nn{\nonumber}
\def\spa#1.#2{\left\langle#1\,#2\right\rangle}
\def\spb#1.#2{\left[#1\,#2\right]}
\def\spaa#1.#2.#3{\langle\mskip-1mu{#1} 
                  | #2 | {#3}\mskip-1mu\rangle}
\def\spbb#1.#2.#3{[\mskip-1mu{#1}
                  | #2 | {#3}\mskip-1mu]}
\def\spab#1.#2.#3{\langle\mskip-1mu{#1} 
                  | #2 | {#3}\mskip-1mu]}
\def\spba#1.#2.#3{[\mskip-1mu{#1} 
                  | #2 | {#3}\mskip-1mu\rangle}
\def\spaba#1.#2.#3.#4{\langle\mskip-1mu{#1} 
                  | #2 | #3 | {#4}\mskip-1mu\rangle}

\def\bentarrow{\:\raisebox{1.3ex}{\rlap{$\vert$}}\!\rightarrow}
\def\bothdk#1#2#3#4#5{
\begin{array}{r c l}
#1 & \rightarrow & #2#3 \\
 & & \:\raisebox{1.3ex}{\rlap{$\vert$}}\raisebox{-0.5ex}{$\vert$}
\phantom{#2}\!\bentarrow #4 \\
 & & \bentarrow #5
\end{array}
}

\title{$W^+W^-$ + jet: compact analytic results}

\ShortTitle{$W^+W^-$ + jet: compact analytic results}

\author{John Campbell\\
        Fermilab\\ 
        E-mail: \email{johnmc@fnal.gov}}

\author{David Miller\\
        University of Glasgow\\ 
        E-mail: \email{d.miller@physics.gla.ac.uk}}

\author{\speaker{Tania Robens}%
         \\
        IKTP, TU Dresden\\
        E-mail: \email{tania.robens@tu-dresden.de}}

\abstract{
In the second run of the LHC, which started in April 2015, an accurate
understanding of Standard Model processes is more crucial than ever. Processes
including electroweak gauge bosons serve as standard candles for SM
measurements, and equally constitute important background for BSM searches. We
here present the NLO QCD virtual contributions to $W^+ W^-$ + jet in an analytic
format obtained through unitarity methods and show results for the full process
using an implementation into the Monte Carlo event generator MCFM.
Phenomenologically, we investigate total as well as differential cross sections
for the LHC with 14~\TeV~ center-of-mass energy, as well as a future 100~\TeV~
proton-proton machine. In the format presented here, the one--loop virtual
contributions also serve as important ingredients in the calculation of
$W^+ W^-$ pair production at NNLO.
\vskip1pt 
\hspace*{\fill} FERMILAB-CONF-16-006-T
}

\FullConference{12th International Symposium on Radiative Corrections (Radcor 2015) and LoopFest XIV (Radiative Corrections for the LHC and Future Colliders)\\
		15-19 June, 2015\\
		UCLA Department of Physics $\&$ Astronomy Los Angeles, USA}

\begin{document}
\bibliographystyle{hunsrt}
\input{yrep}

\section*{Acknowledgements}
T.R. would like to thank the Fermilab Theory Group for their repeated hospitality
while this work was completed. DJM is supported by the UK Science and Technology Facilities Council (STFC) under grant ST/L000446/1.  This research is supported by the US DOE under contract DE-AC02-07CH11359.
\bibliography{yrep}
\end{document}

%% file: yrep.tex
%

%%%%%%%%%%%%%%%%%%%%%%%%%%%%%%%%%%%%%%%%%%%%%%%%%%%%%%%%%%%%%%%%%%%%%5

%\begin{document}
%\usepackage{graphicx}% Include figure files
%\usepackage{dcolumn}% Align table columns on decimal point
%\usepackage{bm}% bold math

%\nofiles
%\bibliographystyle{unsrt}

\section{Overview}
\begin{figure}
\begin{center}
\includegraphics[scale=0.3]{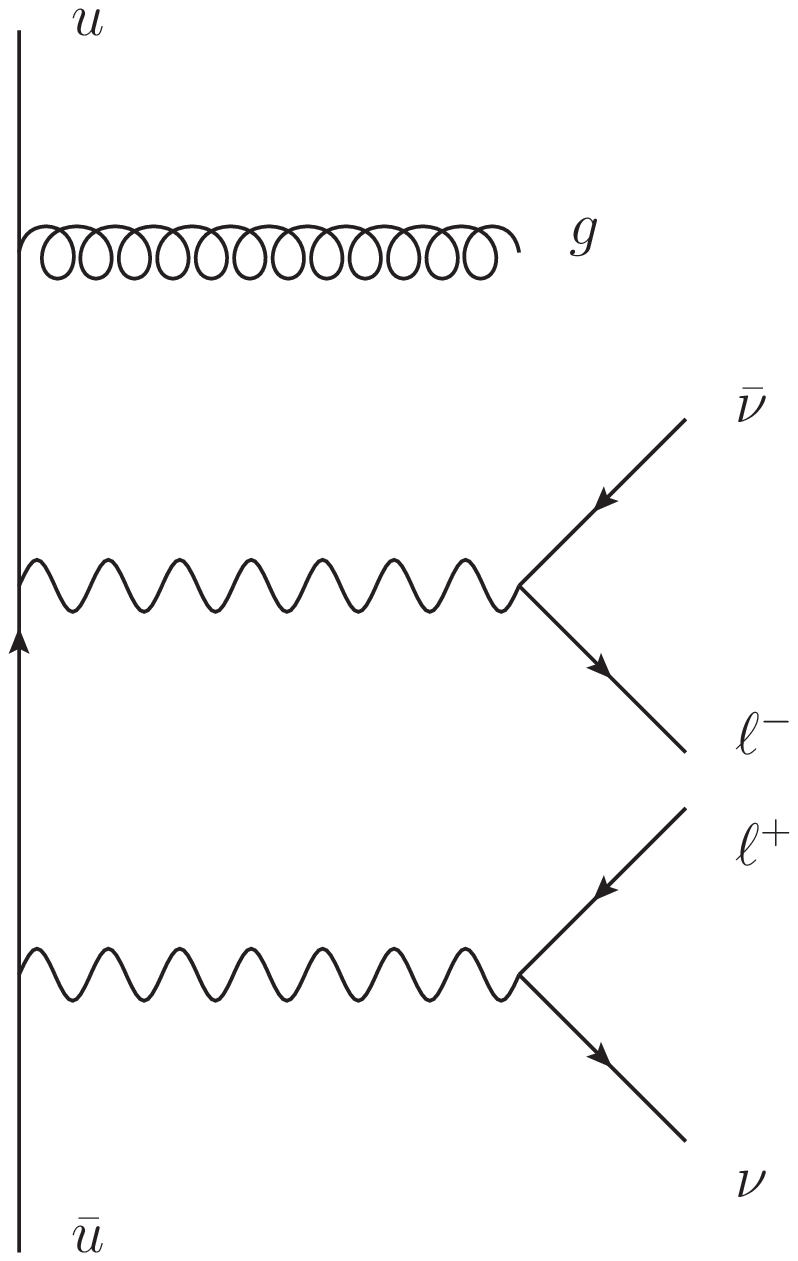} \hspace*{1.5cm}
\includegraphics[scale=0.3]{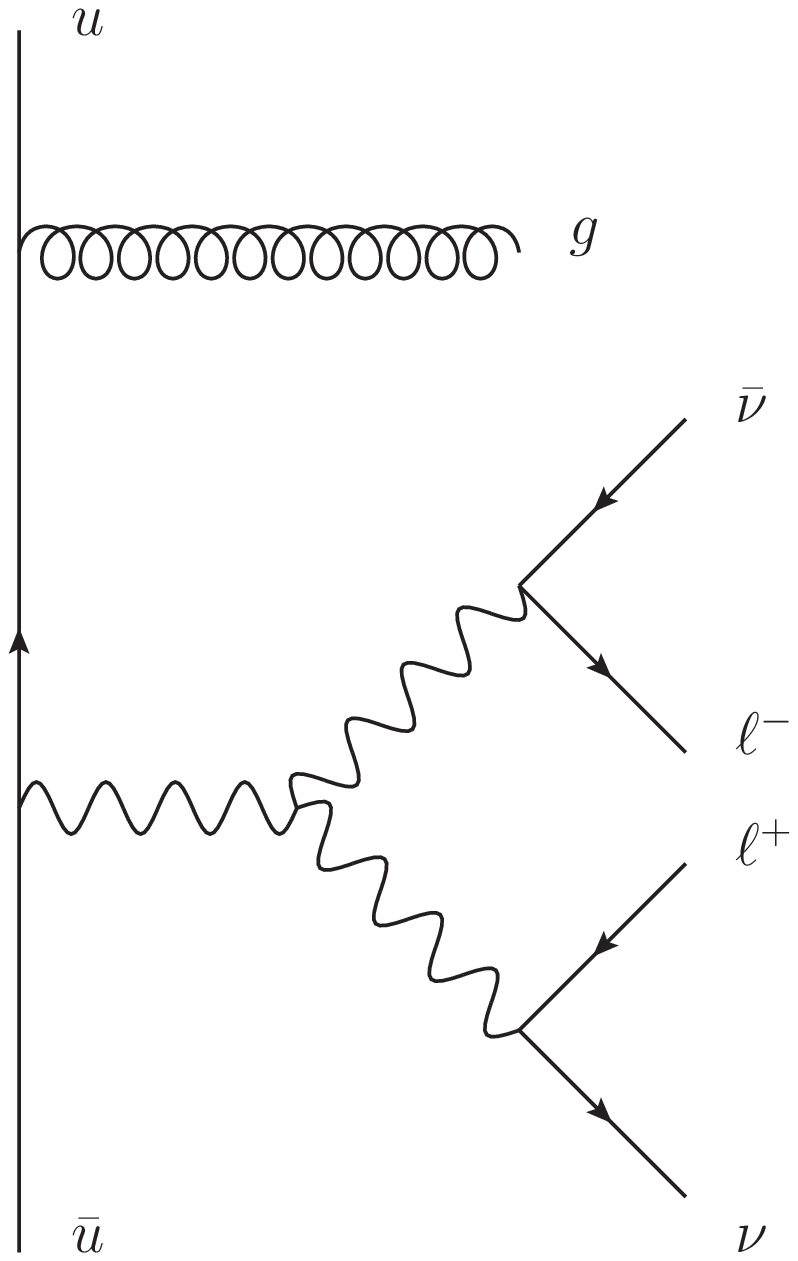} \\ ~~ \\
(a) \hspace*{5cm} (b)
\end{center}
\caption{Sample diagrams entering the calculation of the leading order amplitude
for the $WW+$jet process, corresponding to (a) $W$ emission from the quark line and (b) emission
from an intermediate $Z$ boson or photon.\label{fig:LOdiags}}
\end{figure}

We here consider the hadronic production of $W$ pairs in association with a single
jet at next-to-leading order (NLO) in QCD at a hadron collider with a center-of-mass energy of 14 and 100 \TeV, respectively. The $W$ bosons decay leptonically, with all spin correlations included.  At tree level
this process corresponds to the partonic reaction,
\begin{equation}
\label{WWjetprocess}
\bothdk{q+\bar q}{W^+ +}{W^-+g}{\mu^-+\nu_\mu}{\nu_e+e^+}
\end{equation}
with all possible crossings of the partons between initial and final states. Tree level diagrams for this process are shown in Fig.~\ref{fig:LOdiags}. 

\begin{figure}
\begin{center}
\includegraphics[scale=0.3]{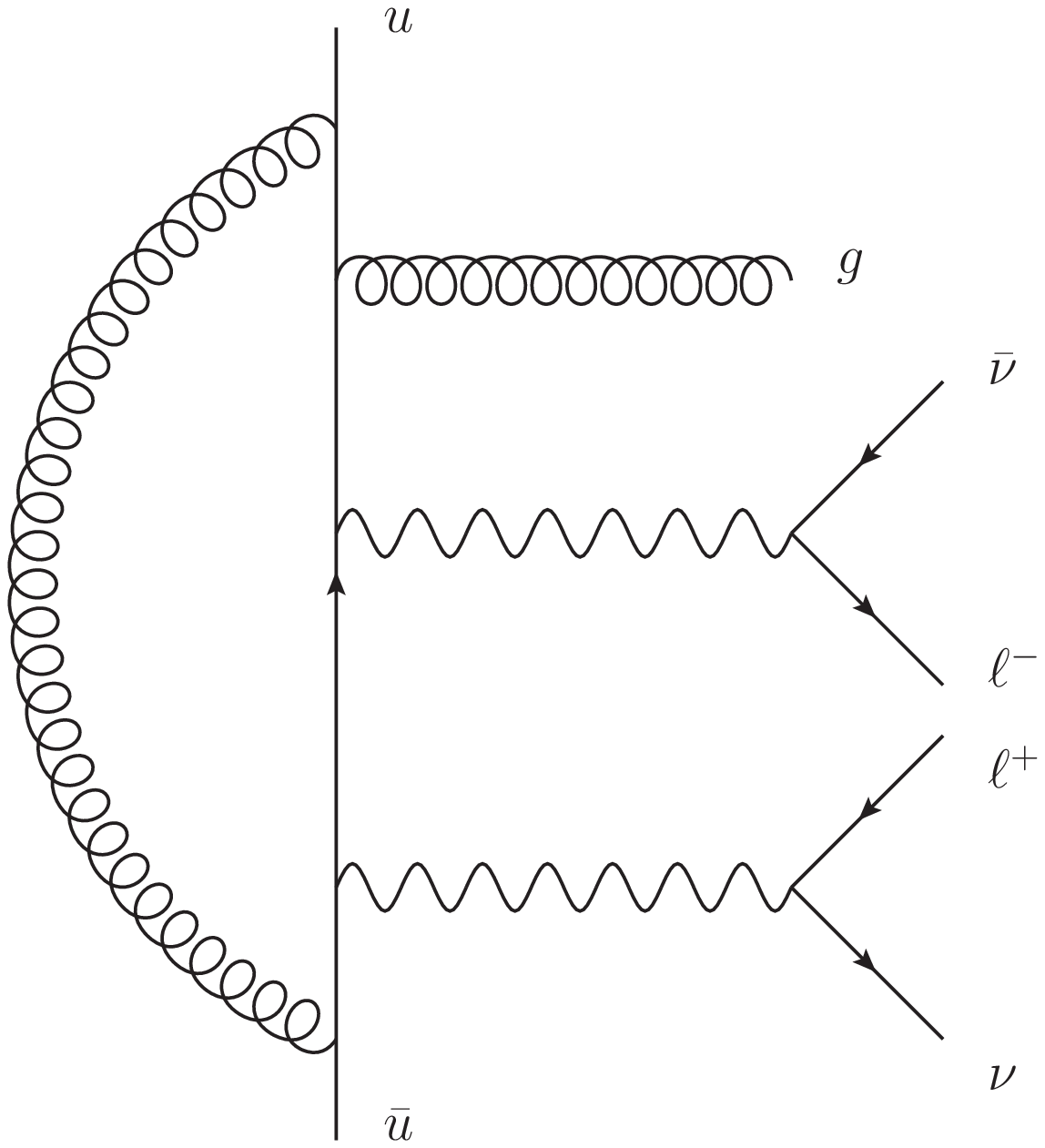} \hspace*{1.5cm}
\includegraphics[scale=0.3]{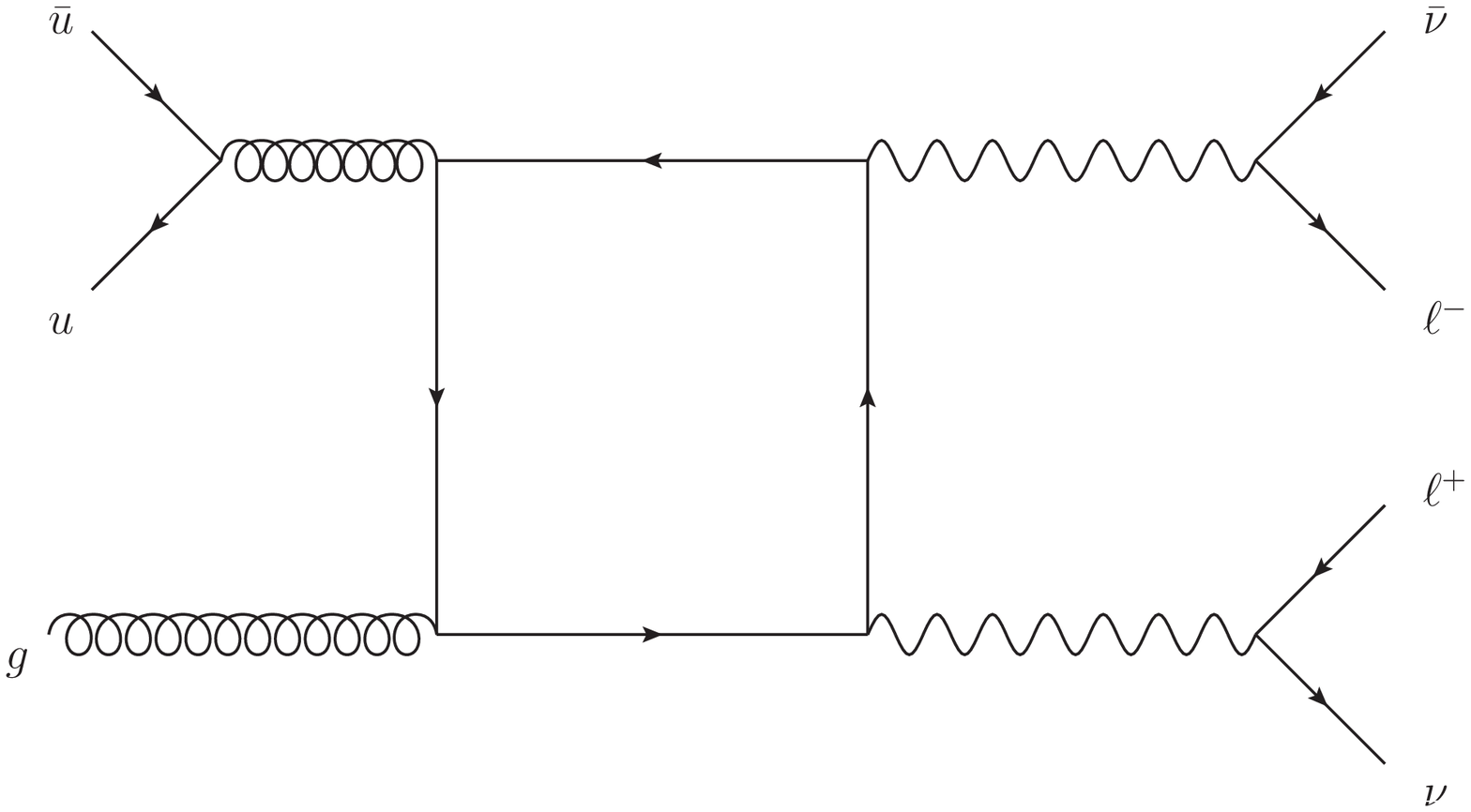}
\end{center}
\caption{Sample diagrams entering the calculation of the one-loop amplitude
for the $WW+$jet process.  The one-loop diagrams can be categorized according
to whether a gluon dresses a leading-order amplitude (left), or whether the
diagram includes a closed fermion loop (right).\label{fig:loopdiags}}
\end{figure}

At next-to-leading order we must include the emission of an additional parton, either as a virtual particle
to form a loop amplitude, or as a real external particle.   Sample diagrams for virtual NLO contributions
are shown in Fig.~\ref{fig:loopdiags}; in addition, one-loop corrections to Fig.~\ref{fig:LOdiags} (b)
must be included. All results presented in the following have been obtained using the calculation of
Ref.~\cite{Campbell:2015hya}, where virtual corrections have been obtained using generalized unitarity
methods \cite{Britto:2004nc,Britto:2005ha,Britto:2006sj,Forde:2007mi,Mastrolia:2009dr,Badger:2008cm} as follows: Each amplitude is decomposed in terms of the usual one-loop basis 
\begin{eqnarray}
\mathcal{A}(\{p_i\}) = \sum_{j} d_{j} I_4^j +  \sum_{j} c_{j} I_3^j +  \sum_{j} b_{j} I_2^j + R \;.
\label{eq:ampdecomp}
\end{eqnarray}
In this equation $I_n^j$ represents a scalar loop integral with $n$ propagators, commonly referred
to as box ($n=4$), triangle ($n=3$) and bubble ($n=2$) integrals.  The integral coefficients
$d_j$, $c_j$ and $b_j$ can be obtained by the application of unitarity cuts in
four dimensions.
The rational remainder term $R$ can be determined using similar cutting rules, after the inclusion
of a fictitious mass for the particles propagating in the loop.  Since the
tree-level on-shell amplitudes that appear in the cutting procedure are quite complex, this procedure
has been performed with the help of the S@M Mathematica package~\cite{Maitre:2007jq}.
The evaluation of the scalar integrals appearing in Eq.~(\ref{eq:ampdecomp}) has been performed using
the QCDLoop Fortran library~\cite{Ellis:2007qk}.

The combination of the virtual contributions with born and real emission diagrams has been implemented
using MCFM~\cite{Campbell:1999ah,Campbell:2015qma}. Note that we do not include the effects of
any third-generation quarks, either as external particles or in internal loops.
\section{Coefficients}
 We consider all particles
outgoing and consider the process,
\beq\label{eq:treeproc}
0 \to q^-(p_1) + {\bar q}^+(p_2) + \ell^-(p_3) + \bar \ell^+(p_4) + \ell^-(p_5) + \bar\ell^+(p_6) + g^+(p_7) \;,
\eeq
 Tree-level amplitudes for this process
have been presented in detail in Refs.~\cite{Dixon:1998py, Campbell:2007ev}, whose notation we follow
closely. 

As a representative box integral coefficient we choose the one
corresponding to the basis integral
$I_4 \leftb s_{56}, s_{34},0, s_{17};s_{127},s_{234}\rightb$.
We here show the leading color integral coefficient, which receives a pre-factor
of $N_c$.  It can be written as, 
\begin{eqnarray}
\lefteqn{d\leftb s_{56}, s_{34},0, s_{17};s_{127},s_{234}\rightb = 
\frac{1}{s_{34}-m_W^2}\,\frac{1}{s_{56}-m_W^2}\frac{\langle  1 2\rangle^2 \,\langle 2|P|2 ]}{2 \, \langle  2 7 \rangle\,\langle 1 7 \rangle} \times} \nn\\
&& \leftb  [{4} {2}]-\frac{\langle {2}|P|{4}] }{D}\rightb \leftb  \langle
   {3}|{2+4}|{6}] -\frac{ \langle {2} {3}\rangle\langle {2}|P|{6}] }{D} \rightb\,\leftb \frac{[{7} {1}] \langle {1} {5}\rangle  }{\langle {2}|P|{7}]}+\frac{ \langle {2} {5}\rangle}{D} \rightb 
\end{eqnarray}
where the compound momentum $P$ and denominator factor $D$ are defined by,
\begin{eqnarray}
P&=&s_{17}\,p_{34}+s_{234}p_{17},\, \qquad
D\,=\,\langle  2|( 3+{ 4})\,( { 1}+ {7})| 2\rangle.
\end{eqnarray}
The factors of $D$ can be put into a more familiar form by relating them to the product
$D D^\star$, where the complex conjugate of $D$ is simply given by $D^\star=[ 2|( 3+{ 4})\,( { 1}+ {7})| 2 ]$.
The product can be written as a trace of gamma matrices that evaluates to,
\begin{equation}
D D^\star = 
  4 s_{34} (p_2 \cdot p_{17} )^2 
+ 4 s_{17} (p_2 \cdot p_{34} )^2 
- 8 (p_2 \cdot p_{17}) (p_2 \cdot p_{34}) (p_{17} \cdot p_{34}) \;.
\label{eq:3boxgram}
\end{equation}
This is just the Gram determinant for this basis integral;  its presence,
when raised to a sufficiently high power, can lead to numerical instability
in phase space regions where it is very small. To avoid any such issues we
veto phase regions where cancellations between the terms in
Eq.~(\ref{eq:3boxgram}) (and equivalent expressions for the other
box integrals) occur at the level of $10^{-6}$ or more.
In our studies this occurs only very rarely, in about one in a million
events, so that the effects of such a veto are tiny
compared to the anticipated level of precision.
\section{Total cross sections}
\begin{figure}
\begin{center}
\includegraphics[scale=0.45,angle=-90]{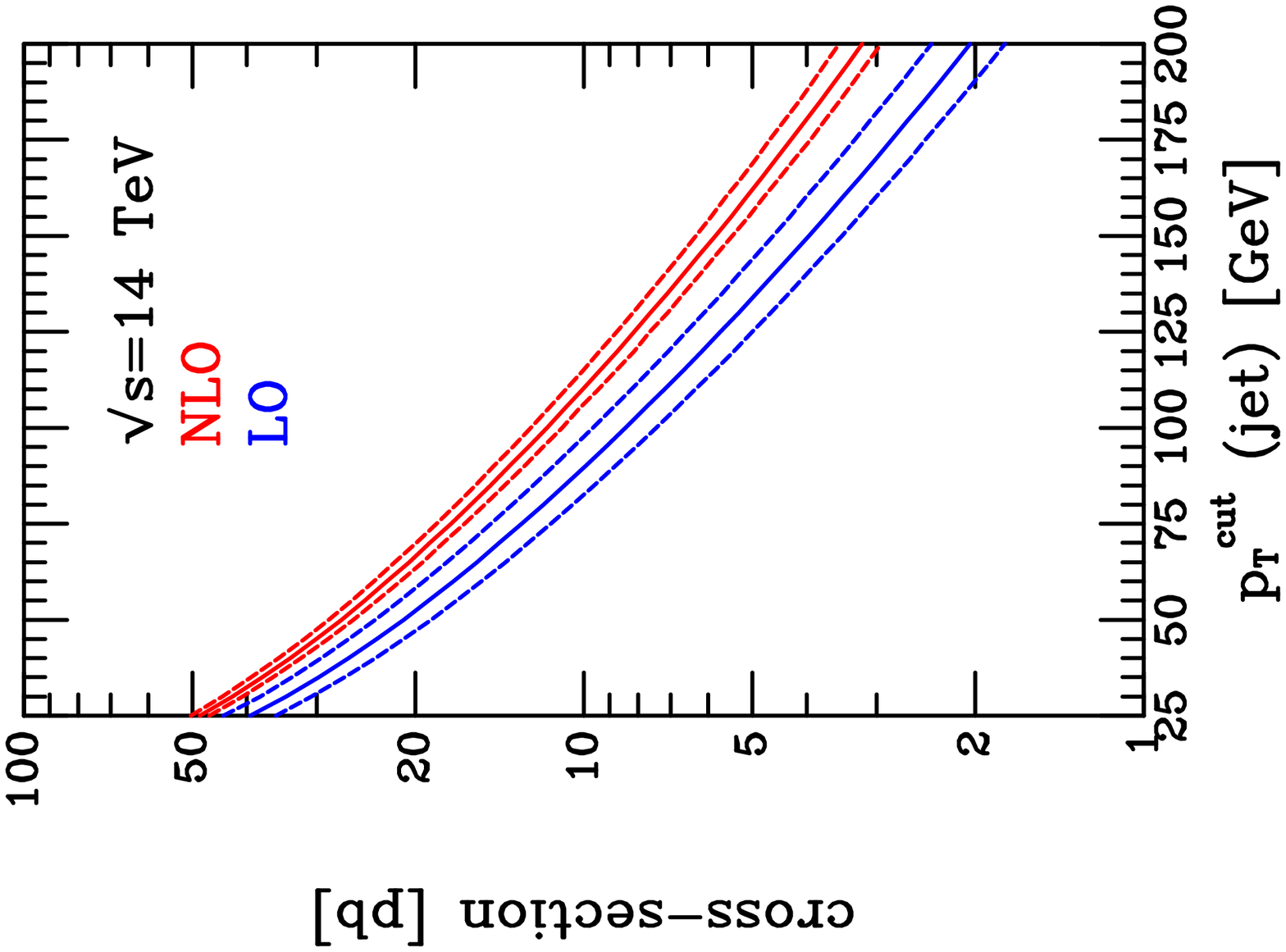} \hspace*{0.5cm}
\includegraphics[scale=0.45,angle=-90]{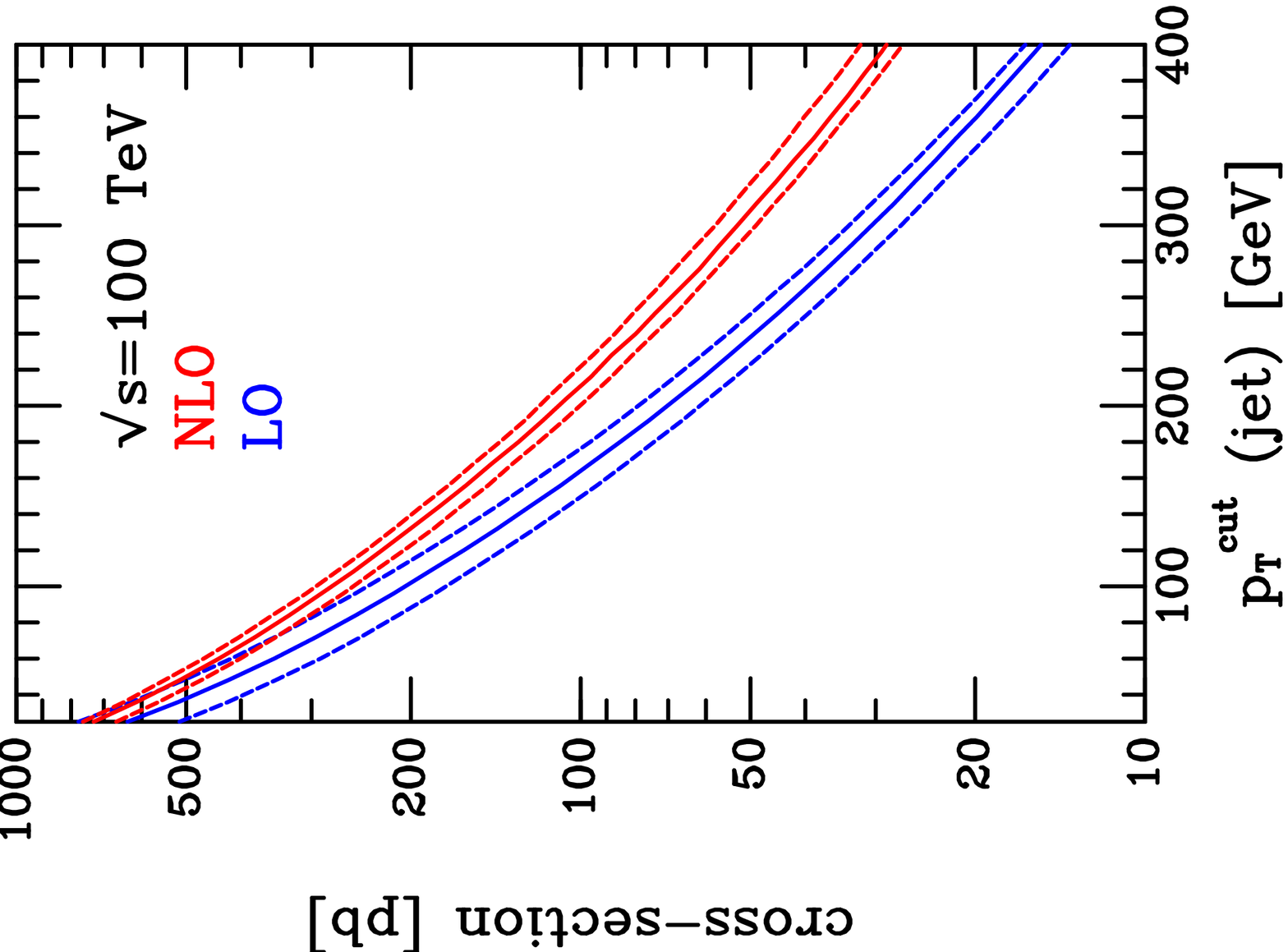}
\end{center}
\caption{Cross-sections at $\sqrt s = 14$~TeV (left) and $100$~TeV (right),
as a function of the transverse momentum
cut on the jet.  The prediction at each order is shown as a solid line,
with the dotted lines indicating the scale uncertainty corresponding to a factor of two variation
about the central scale.
\label{fig:ptjet}}
\end{figure}

\begin{table}
\begin{center}
\begin{tabular}{|c|c|c|c|}
\hline
$m_W$               & 80.385 GeV           & $\Gamma_W$ & 2.085 GeV \\
$m_Z$               & 91.1876 GeV          & $\Gamma_Z$ & 2.4952 GeV \\
$e^2$               & 0.095032             & $g_W^2$    & 0.42635 \\ 
$\sin^2\theta_W$    & $0.22290$            & $G_F$      & $0.116638\times10^{-4}$ \\
\hline
\end{tabular}
\caption{The values of the mass, width and electroweak parameters used to produce
the results in this paper.
\label{parameters}}
\end{center}
\end{table}
The results presented in this section have been obtained using the parameters shown in
Table~\ref{parameters}. 
In calculations of LO quantities we employ the CTEQ6L1 PDF set~\cite{Pumplin:2002vw},
while at NLO we use CT10~\cite{Lai:2010vv}.  The renormalization and factorization scales are
usually chosen to be the same, $\mu_R = \mu_F = \mu$, with our default scale choice
$\mu = \mu_0$ given by,
\begin{equation}
\mu_0 \equiv \frac{H_T}{2} = \frac{1}{2} \sum_i p_{\perp}^i \;.
\end{equation}
The sum over the index $i$ runs over all final state leptons and partons.
Jets are defined using the anti-$k_T$ algorithm with separation parameter
$R=0.5$ and must satisfy,
\begin{equation}
p_{\perp}^\text{jet} > p_{\perp, \text{cut}}^{\text{jet}} \;, \qquad
|\eta^\text{jet}| <4.5 \;.
\label{eq:jetcuts}
\end{equation}
The cross-sections predicted at LO and NLO are shown in Fig.~\ref{fig:ptjet}, as a function
of $p_{\perp, \text{cut}}^{\text{jet}}$ and for values as large as $400$~GeV at the
$100$~TeV machine.
The theoretical uncertainty band is computed by using
a series of scale variations about the central choice $\mu_0$.
The uncertainty corresponds to scale variations of
%\begin{equation}
$\left\{ \mu_R, \mu_F \right\} =
 \left\{ 2 \mu_0, 2\mu_0 \right\} ,
 \left\{ \mu_0/2, \mu_0/2 \right\} $
% \left\{ 2 \mu_0, \mu_0/2 \right\} ,
% \left\{ \mu_0/2, 2\mu_0 \right\} \;.
%\label{eq:scaleuncert} 
%\end{equation}
for 14 TeV and 
%\begin{equation}
$\left\{ \mu_R, \mu_F \right\} =
% \left\{ 2 \mu_0, 2\mu_0 \right\} ,
% \left\{ \mu_0/2, \mu_0/2 \right\} ,
 \left\{ 2 \mu_0, \mu_0/2 \right\} ,
 \left\{ \mu_0/2, 2\mu_0 \right\} $
%\label{eq:scaleuncert} 
%\end{equation}
for 100 TeV.  
The cross-sections at NLO are significantly larger than those at LO and, in
general, the uncertainty bands do not overlap.   At 100 TeV the cross-sections
are about an order of magnitude larger than at 14 TeV. 

As useful operating points, we use $p_{\perp, \text{cut}}^{\text{jet}}=25~\mbox{GeV}$ at both
collider energies and also choose to study the additional case
$p_{\perp, \text{cut}}^{\text{jet}}=300~\mbox{GeV}$ at $100$~TeV,
which we will label 100 \TeV * in the following.
The cross-sections for $WW$+jet production at these colliders, under the basic jet cuts
of Eq.~(\ref{eq:jetcuts}), are collated in Table~\ref{xsecs}
\footnote{Note that there is a minor typographical error in
Ref.~\cite{Campbell:2015hya} in the relative uncertainty due to scale variations for the LO cross
section at 100 \TeV, which we have corrected here.}.  Note that the effect of the decays of the $W$
bosons is not included. At the 100 TeV machine, the jet cut of 300~GeV has been chosen so that the
cross section is similar in size to the 14 TeV cross section, as can be seen from
Table~\ref{xsecs}.  This cut provides a useful benchmark in a different kinematic regime
that may be more appropriate at that collider energy.
\renewcommand{\baselinestretch}{1.5}
\begin{table}
\begin{center}
\begin{tabular}{|r|r|c|c|}
\hline
$\sqrt s$~~~~ & $p_{\perp, \text{cut}}^{\text{jet}}$ & $\sigma_{LO}$~[pb] & $\sigma_{NLO}$~[pb] \\
\hline
%$13$~TeV      & $34.9_{-11.0\%}^{+11.4\%}$  & $42.9_{-3.7\%}^{+3.7\%}$ \\
$14$~TeV      &  25 GeV & $39.5_{-11.0\%}^{+11.7\%} $  & $48.6_{-4.0\%}^{+3.8\%}$ \\
$100$~TeV     &  25 GeV & $648_{-19.3\%}^{+22.3\%}   $  & $740_{-9.3\%}^{+4.5\%} $ \\
$100$~TeV     & 300 GeV & $30.3^{+11.22\%}_{-10.56\%}$&$53.7^{+8.0\%}_{-7.6\%}$ \\
\hline
\end{tabular}
\renewcommand{\baselinestretch}{1.0}
\caption{Cross-sections for the process $p p \to WW$+jet at proton-proton colliders
of various energies, together with estimates of the theoretical uncertainty
from scale variation as described in the text. 
Monte Carlo uncertainties are at most a
single unit in the last digit shown shown in the table.
\label{xsecs}}
\end{center}
\end{table}
\renewcommand{\baselinestretch}{1.0}

\begin{figure}
\begin{center}
\includegraphics[width=0.3\textwidth,angle=-90]{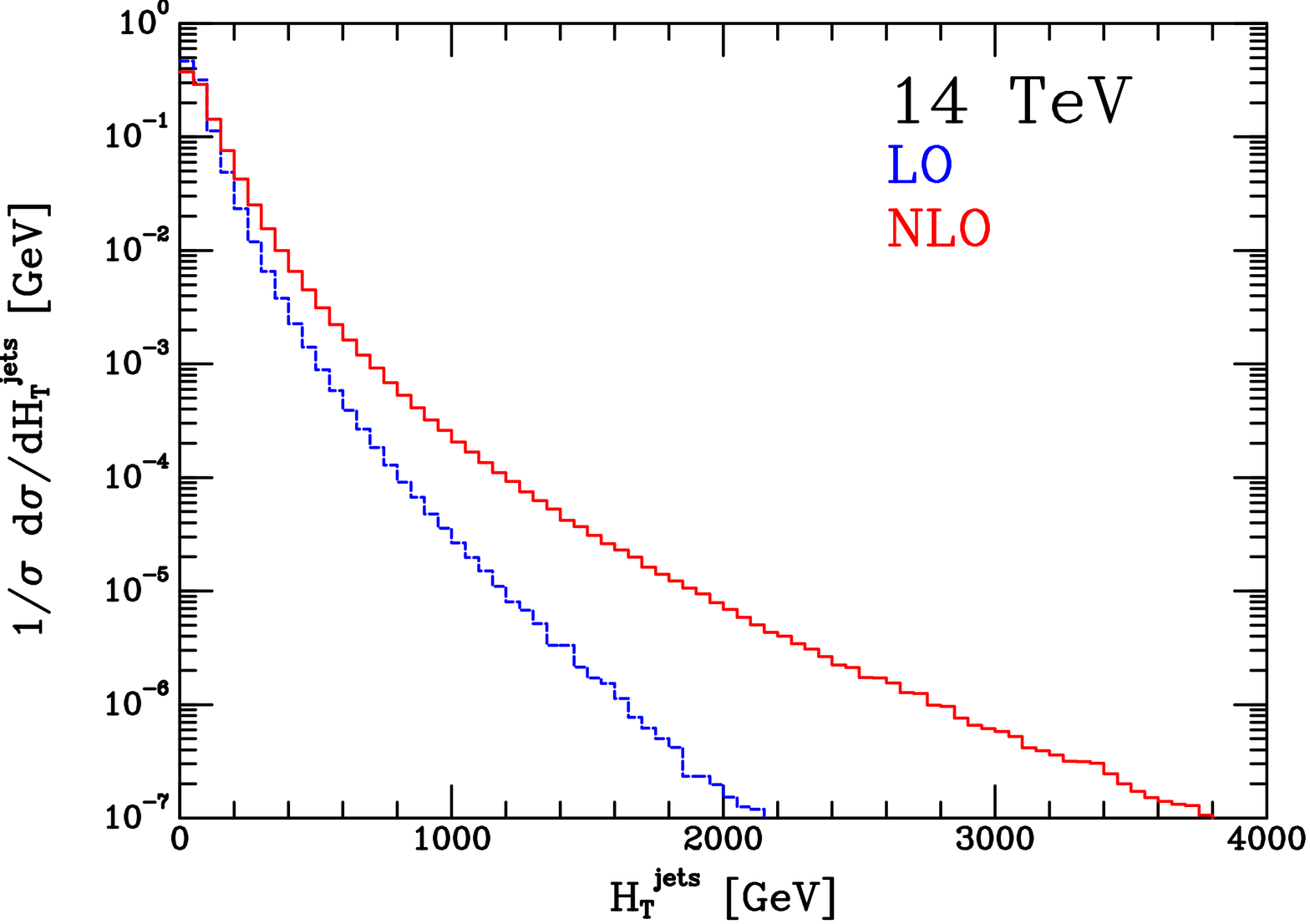} \hspace*{0.5cm}
\includegraphics[width=0.3\textwidth,angle=-90]{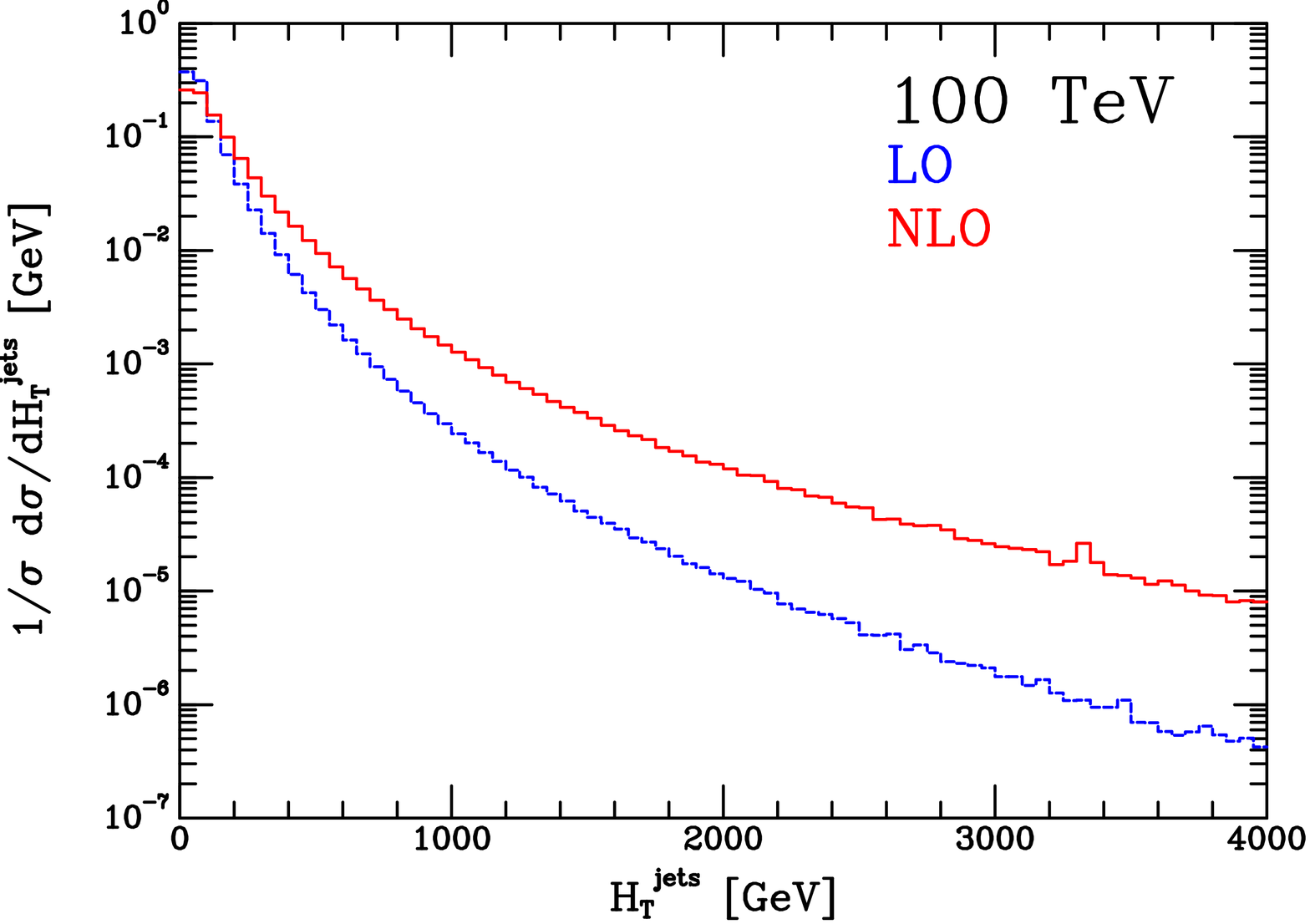}
\end{center}
\caption{The distribution of the observable
$H_T^\text{jets}=\sum_\text{jets} p_\perp^\text{jet}$ at LO and NLO,
for 14 TeV (left) and 100 TeV (right).
\label{fig:HTjets}}
\end{figure}

An interesting feature of the higher order corrections to processes such
as the one at hand is the existence of so-called
``giant K-factors''~\cite{Rubin:2010xp}.  These are due to the existence
of kinematic configurations at NLO that do not exist at LO and that
can be the dominant contribution in certain distributions.  An observable
that exemplifies this effect is $H_T^\text{jets}$, which is defined to
be the scalar sum of all jet transverse momenta in a given event.  At NLO,
real radiation contributions arise in which two hard partons are produced
approximately back-to-back, with the $W^+W^-$ system relatively soft.  Such
configurations are not captured at all by the LO calculations, in which the 
parton and $W^+W^-$ system are necessarily balanced in the transverse plane. 
This results in the by now well-known feature of huge NLO corrections at
large $H_T^\text{jets}$, as shown in Fig.~\ref{fig:HTjets}.
Similar effects are observed at both energies, with NLO predictions at least
an order of magnitude larger than their LO counterparts in the tails of the
distributions.  At $100$~TeV the onset of the giant $K$-factor is a little
slower, but still occurs well before the interesting multi-TeV region.

\section{Differential distributions}
We first consider the case of $14$~TeV LHC running, with a set of cuts
inspired by the ATLAS determination of the spin and parity of the Higgs boson
presented in Ref.~\cite{Aad:2015rwa}.  The $WW$ process constitutes the
largest irreducible background in the $H \to WW^\star$ decay channel and
a cocktail of cuts must be applied in order to access information about
the Higgs boson. The cuts
are summarized in Table~\ref{14TeVcuts}.  These include constraints on
the transverse mass of $(X,E^\text{miss}_T)$ systems, $m_T^X$,
where $X \in (\ell\ell, \ell_1, \ell_2)$, with
$p_{\ell\ell}\,=\,p_{\ell_1}+p_{\ell_2}$.  This quantity is defined
by\footnote{See, for instance, Eq.~(3) of Ref.~\cite{Khachatryan:2015cwa}.},
\begin{equation}
m_T^X\,=\,\sqrt{2\,p_{\perp}^X E^\text{miss}_T\,
 \left( 1-\cos\Delta\Phi(\ora{p}^X_T,\ora{E}_T^\text{miss})  \right)}.
\end{equation}

\begin{center}
\begin{table}
\begin{center}
\begin{tabular}{l|l}
variable& cut \\ \hline
$p_{\perp,j}$&$>$ 25 \GeV\\
$|\eta_j|$&$<4.5$ \\
\hline
$|\eta_\ell|$&$<$ 2.5 \\
$p_{\perp,\ell_1}$&$>$ 22 \GeV \\
$p_{\perp,\ell_2}$&$>$ 15\,\GeV \\
$m_{\ell\ell}$&$\in[10, 80]\,\GeV$ \\
$p_\perp^\text{miss}$&$>$ 20\,\GeV \\
$\Delta\Phi_{\ell\ell}$&$<$ 2.8 \\
$m_T^{\ell\ell}$&$<$ 150 \GeV \\
$\text{max}[m_T^{\ell_{1}},m_T^{\ell_2}]$&$>$ 50 \GeV 
\end{tabular}
\caption{ Cuts applied in the $14$~TeV analysis, corresponding to
the ``full'' set of cuts.  The jet cuts, corresponding to the
first two lines in the table, are the only ones applied for
the ``basic'' cross-section.}
\label{14TeVcuts}
\end{center}
\end{table}
\end{center}

In the results that follow we shall
always consider the decay of each $W$ boson into a single lepton family, i.e.
the Born level quark-antiquark process we consider is the one shown
in Eq.~(\ref{WWjetprocess}).
The cross-sections under these cuts are given in Table~\ref{14TeVresults}.
In order to assess their effect, we also show for comparison the cross
sections obtained using only the jet cuts, i.e. the top two lines
of the cuts in Table~\ref{14TeVcuts}. The table also shows the $K$-factor, defined by $K=\sigma^{\rm{NLO}}/\sigma^{\rm{LO}}$,
which we find is rather insensitive to which set of cuts is applied.
\begin{center}
\begin{table}
\begin{center}
\begin{tabular}{l|l|l|l}
cuts ~~~& $\sigma^{\rm{LO}}$ [fb] ~~& $\sigma^{\rm{NLO}}$ [fb]~~&~~$K$~~ \\ \hline
basic   & 462.0(2)                & 568.4(2)& 1.23 \\
full    & 67.12(4)                & 83.91(5)& 1.25 \\
%spin-2  & 58.21(4)                & 71.32(5)& 1.23 \\
\end{tabular}
\caption{Cross-sections at 14 TeV.
Monte Carlo uncertainties are indicated in
parentheses and are smaller than the per mille level.} 
\label{14TeVresults}
\end{center}
\end{table}
\end{center}

\begin{figure}
\begin{minipage}{0.45\textwidth}
\includegraphics[height=\textwidth,angle=-90]{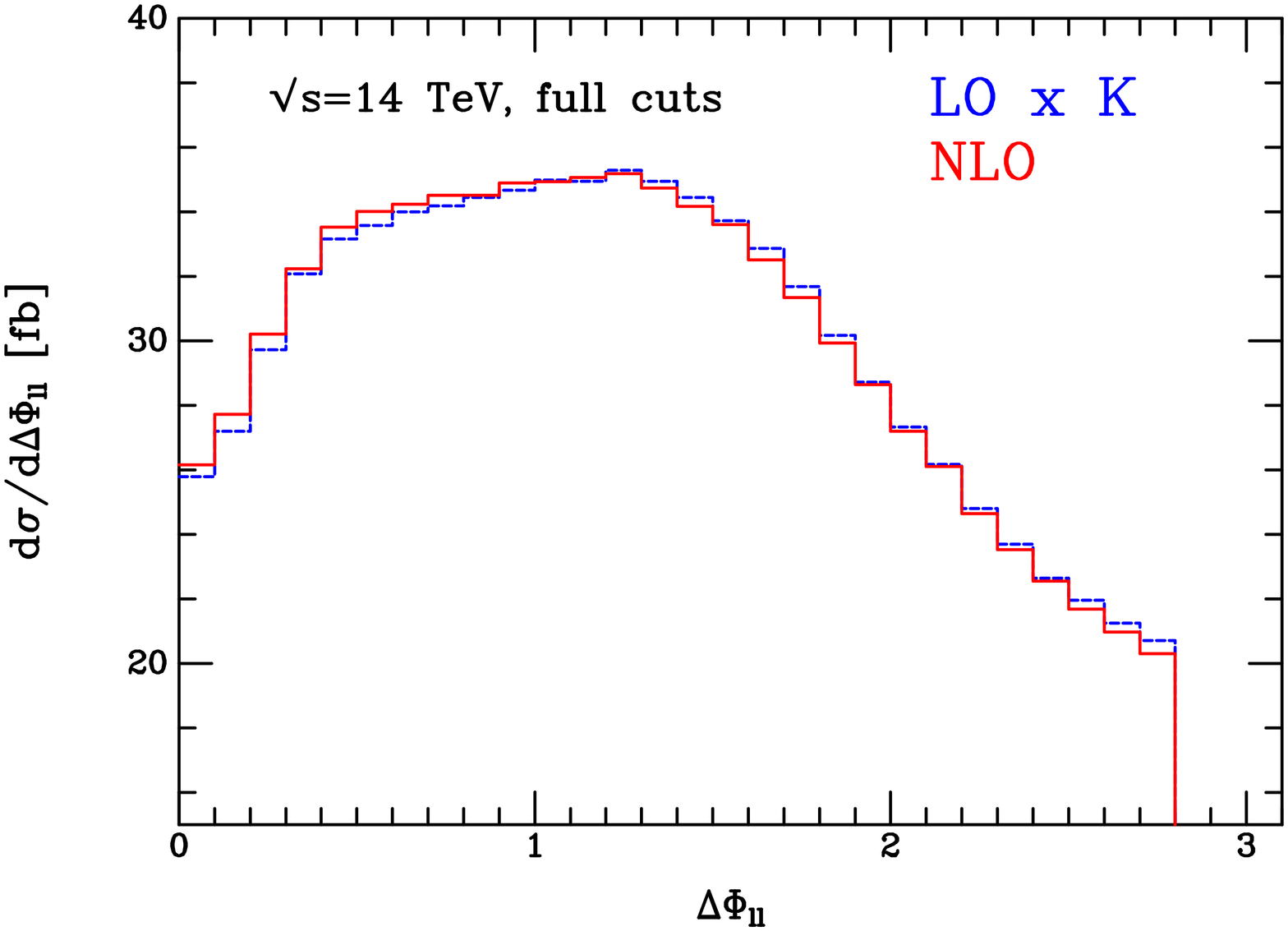}
\end{minipage}
%\hspace{-12mm}
\begin{minipage}{0.45\textwidth}
\includegraphics[height=\textwidth,angle=-90]{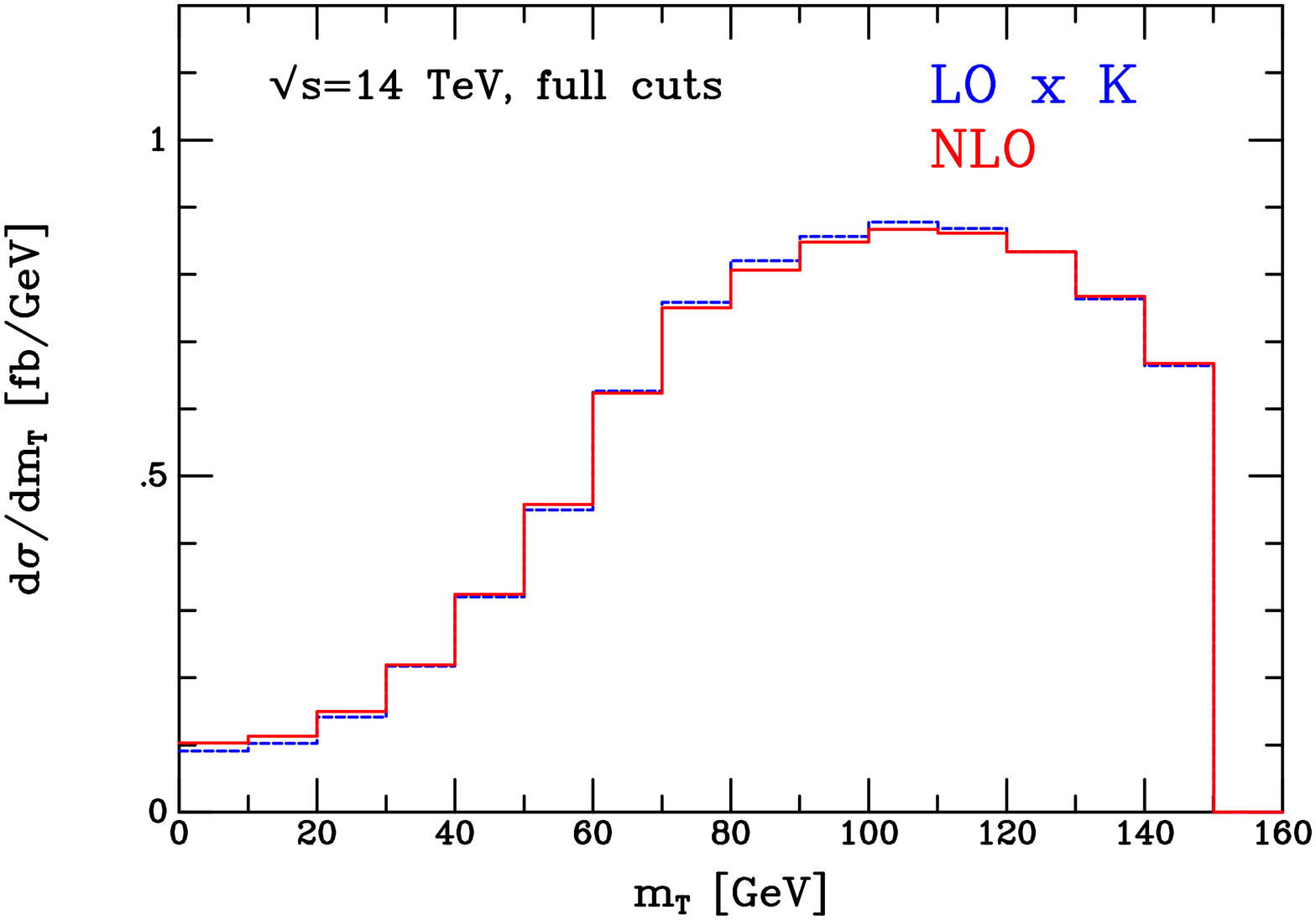}
\end{minipage}\\
\begin{minipage}{0.45\textwidth}
\includegraphics[height=\textwidth,angle=-90]{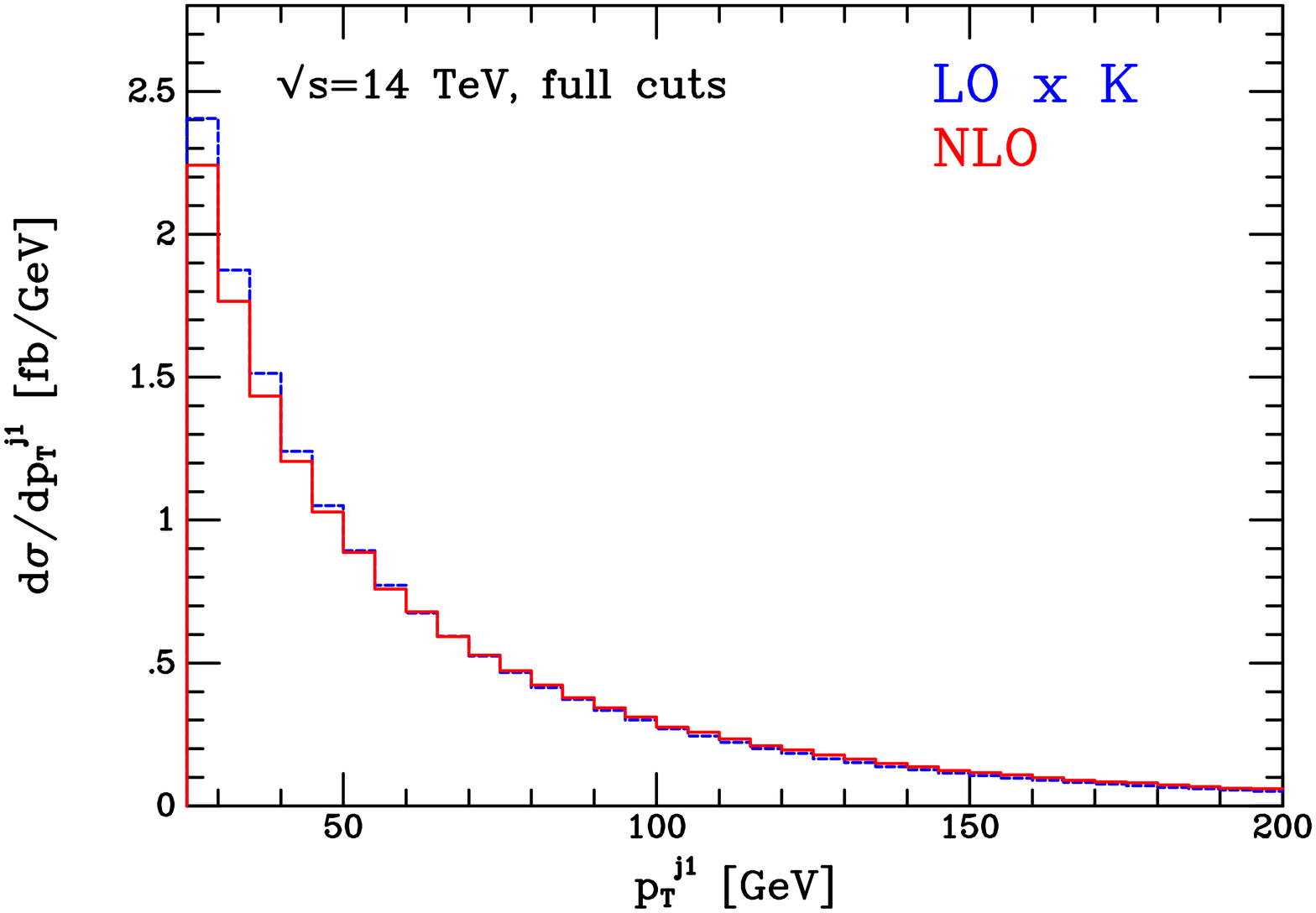}
\end{minipage}
\begin{minipage}{0.45\textwidth}
\includegraphics[height=\textwidth,angle=-90]{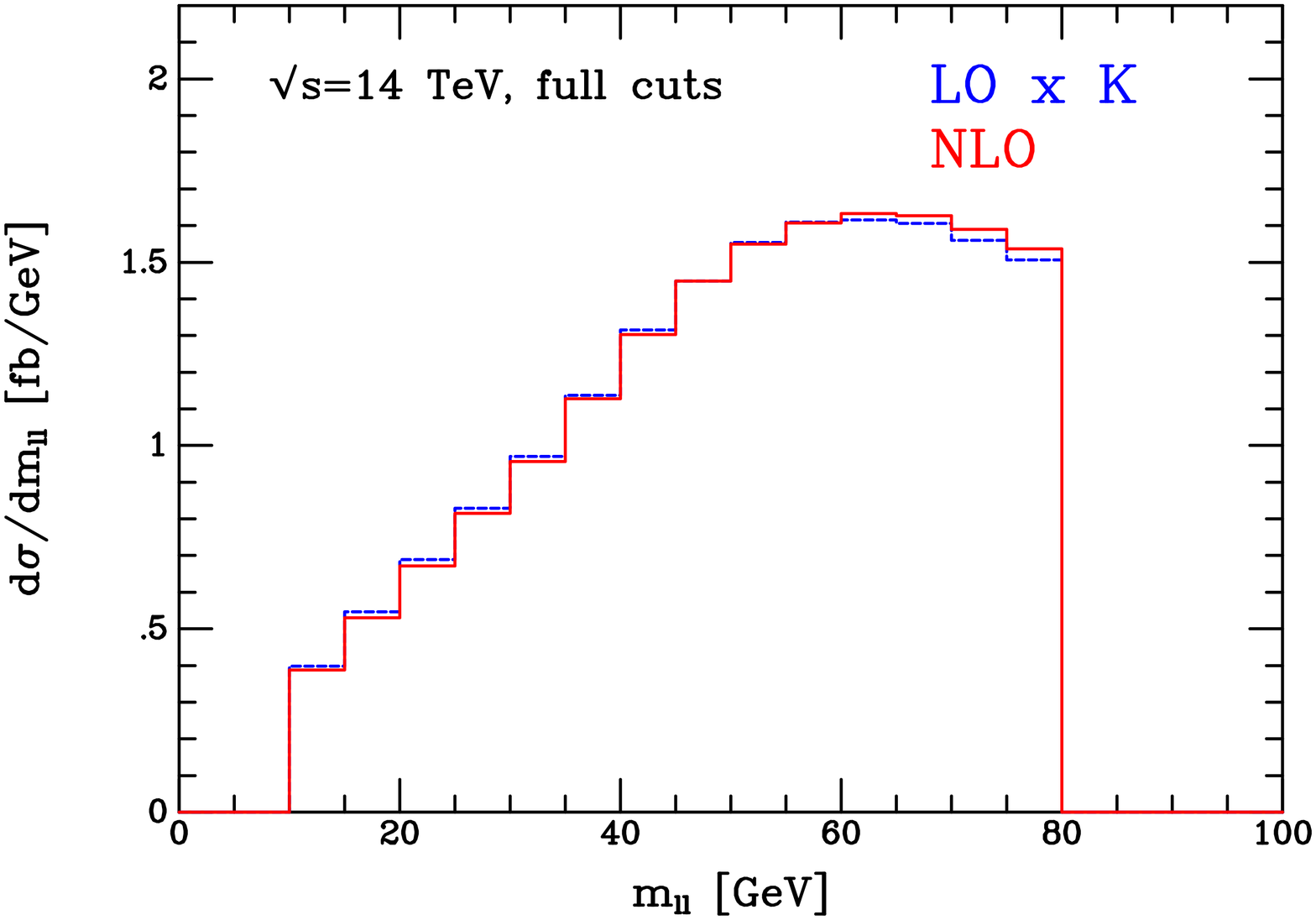}
\end{minipage}
\caption{ Kinematic distributions at 14 \TeV,
using the full set of cuts specified in the text.  The NLO prediction
is shown as the solid (red) histogram, while the dashed (blue) histogram
corresponds to the LO prediction rescaled by the $K$-factor.}
\label{fig:14TeV}
\end{figure}

We now consider differential distributions in $m_T^{\ell\ell}$, $\Delta \Phi_{\ell\ell}$ and $m_{\ell\ell}$
as well as the transverse momentum of the lead jet, $p_{\perp}^{j_1}$.
These quantities are shown in Figure~\ref{fig:14TeV} where, for comparison,
the LO prediction has been rescaled by the $K$-factor from Table~\ref{14TeVresults}.
This indicates that there is very little difference
between the shapes of the distributions at each order, with the exception
of the transverse momentum of the leading jet.  In contrast this does receive
significant corrections, which is expected since additional radiation beyond
a single jet is only present at NLO.

\begin{figure}
\begin{center}
\includegraphics[width=0.33\textwidth,angle=-90]{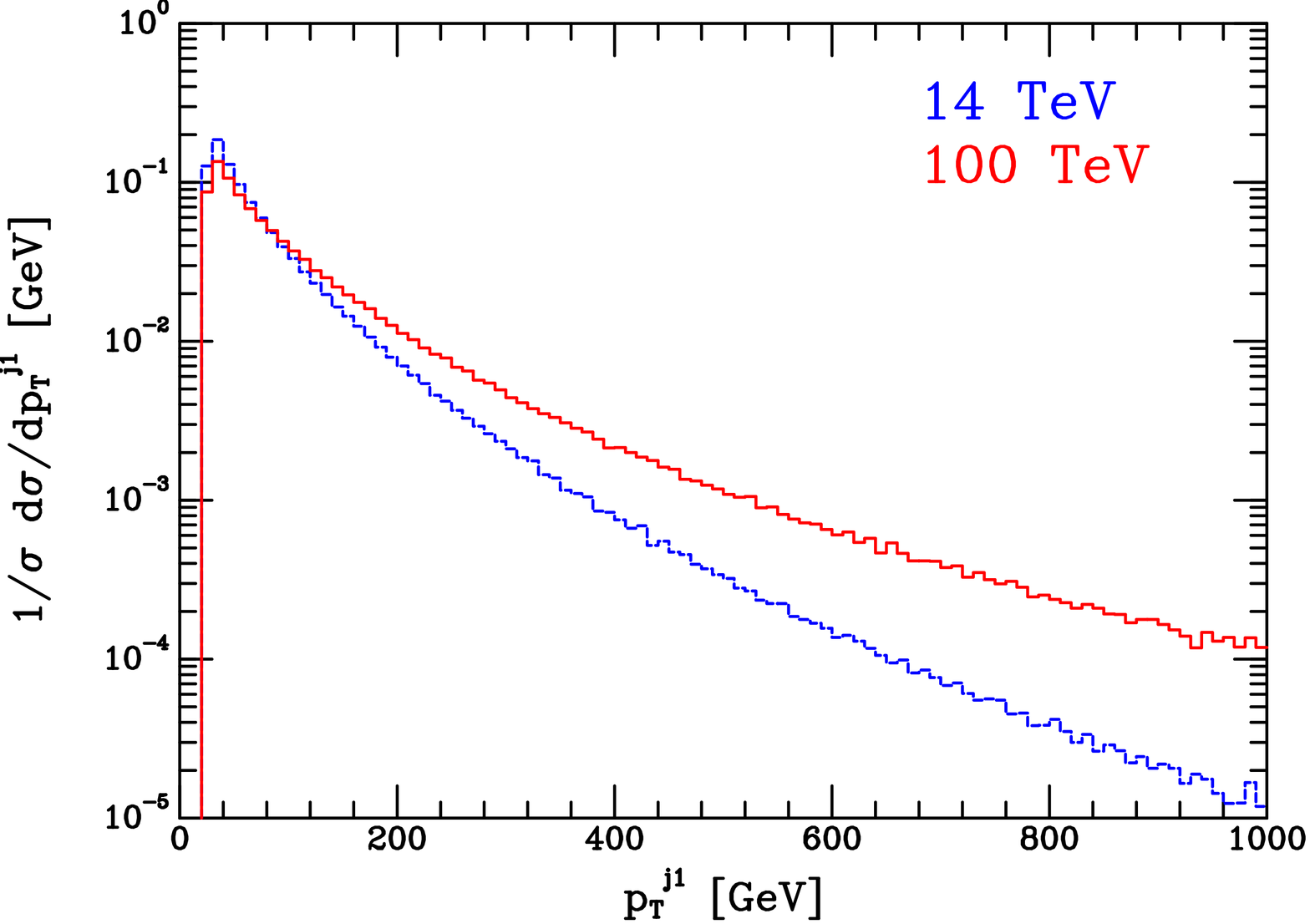} \hspace*{0.5cm}
\includegraphics[width=0.33\textwidth,angle=-90]{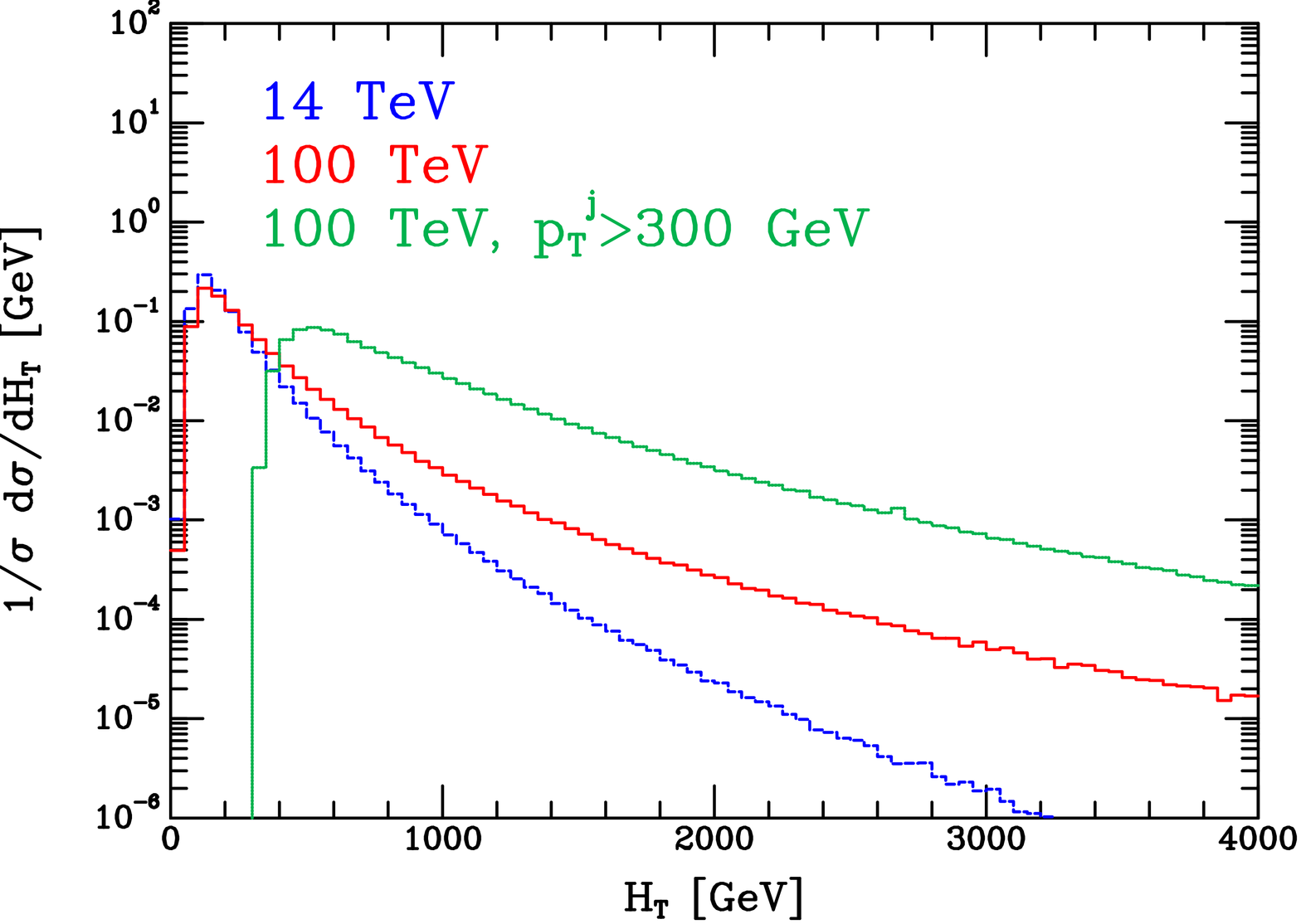}
\end{center}
\caption{NLO $p_{\perp, j}$ (left) and $H_T$ (right) distributions, normalized by the respective
total cross sections, for 14 \TeV (red), 100 \TeV (blue), and 100 \TeV * (green)
\label{fig:comp1}}
\end{figure}

To illustrate some of the key differences between the predictions for $WW$+jet production at 14 and 100 \TeV, we now examine NLO predictions for a number of kinematic distributions and compare their behaviour at different c.o.m energies.
Fig.~\ref{fig:comp1} shows two quantities that characterize the overall nature of this
process, the transverse momentum of the leading jet and the scalar sum of all jet and lepton
transverse momenta, $H_T$ (c.f. $H_T^\text{jets}$ earlier).
All histograms have been normalized to the total NLO cross-sections
given earlier, in order to better compare their shapes.  At $100$~TeV the leading jet is
significantly harder than at $14$~TeV.  The $H_T$ distribution is also harder at $100$~TeV with,
of course, a significant shift in the peak once the jet cut is
raised.~\footnote{This variable is also frequently used as a cut variable
in searches for physics beyond the SM, for example in
Refs.~\cite{Aad:2014gka,Aad:2015wqa}, where cuts are placed in the
 range $\sim 0.6$--$2\TeV$ depending on the
 details of the search strategy.}

\begin{figure}
\begin{center}
\includegraphics[width=0.33\textwidth,angle=-90]{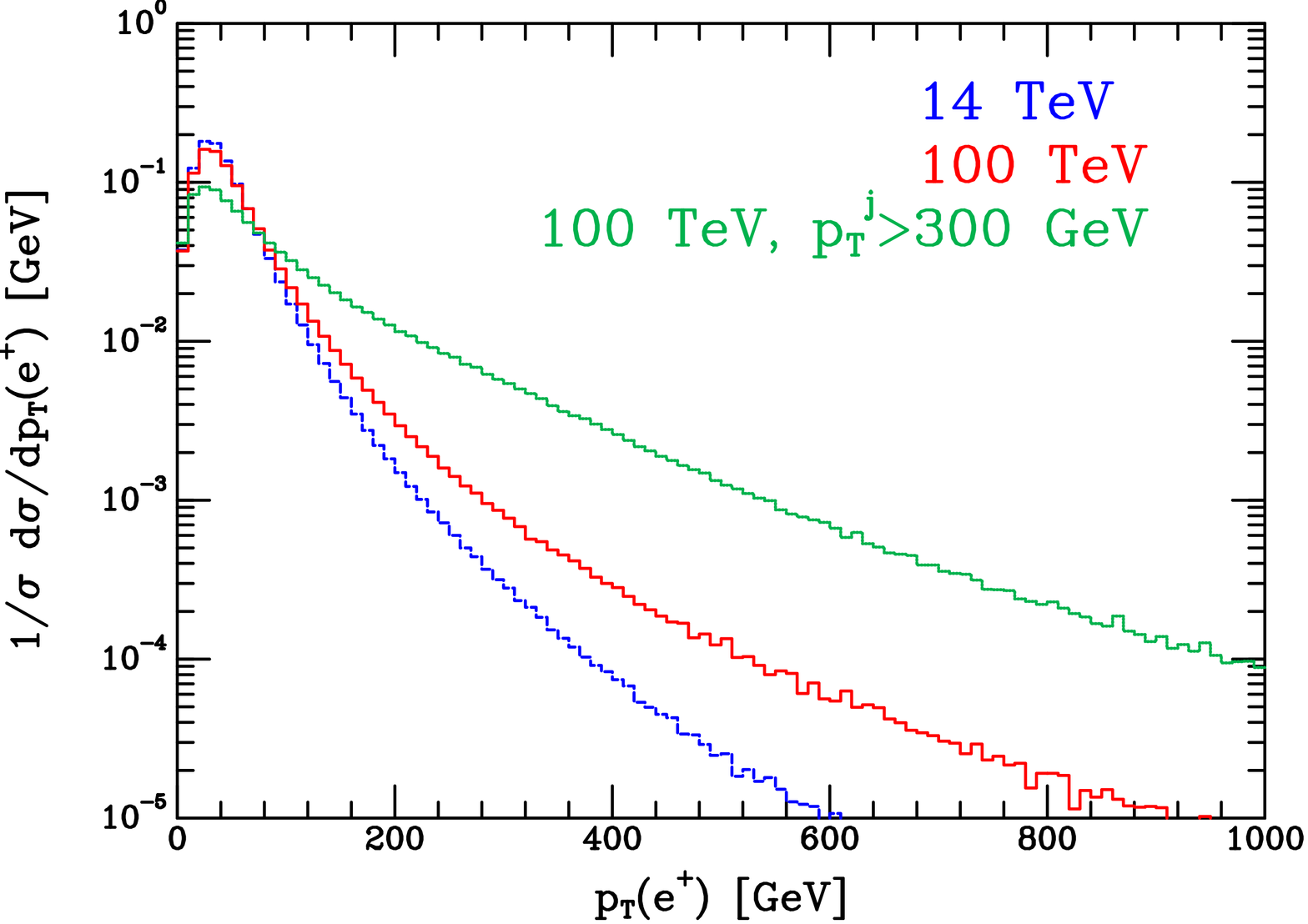} \hspace*{0.5cm}
\includegraphics[width=0.33\textwidth,angle=-90]{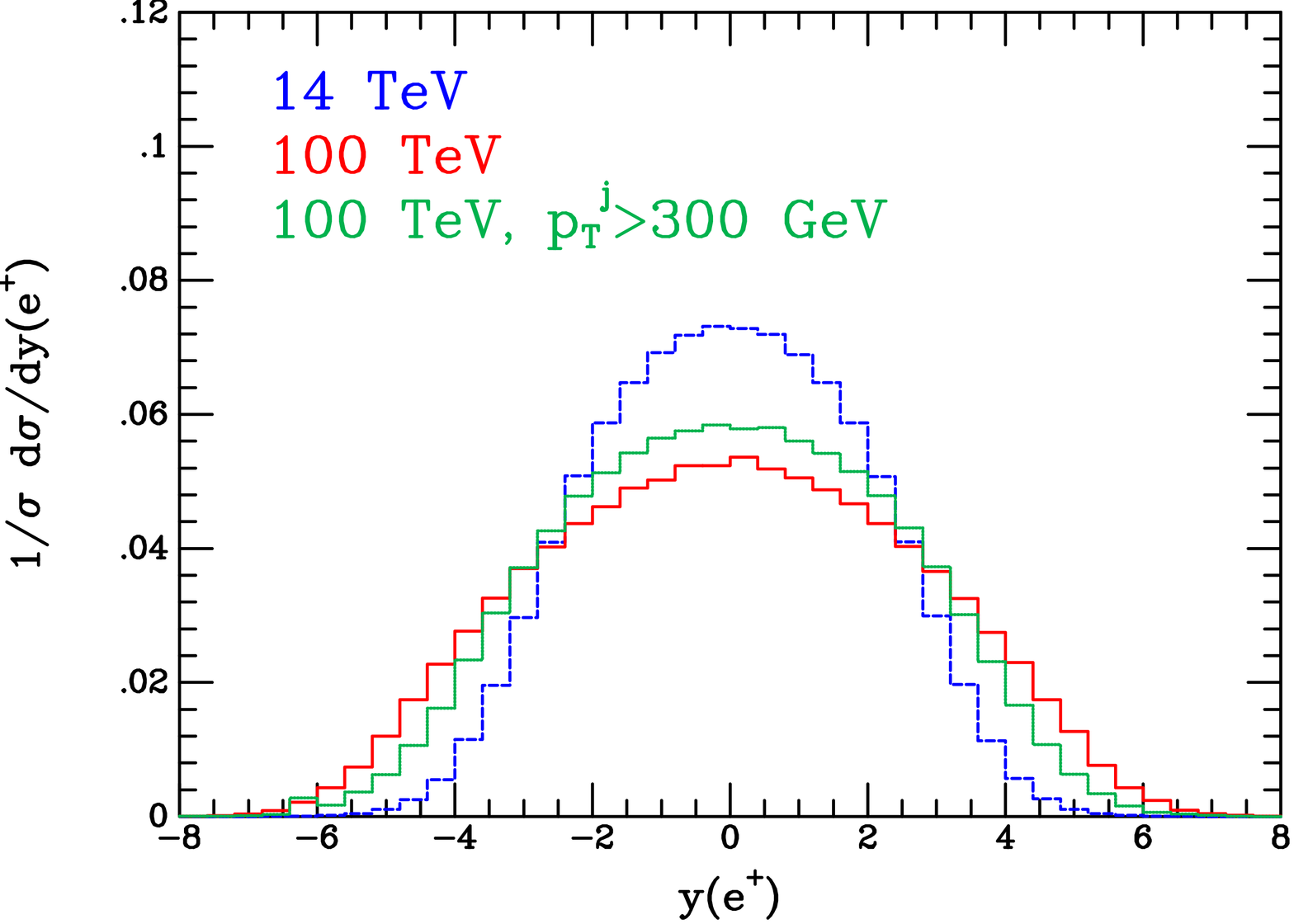}
\end{center}
\caption{NLO  $p_{\perp, \ell}$ (left) and $\eta_\ell$ (right)
distributions, normalized by the respective total cross sections, for 14 \TeV (red),
100 \TeV (blue), and 100 \TeV * (green) 
\label{fig:comp2}}
\end{figure}

\begin{figure}
\begin{center}
\includegraphics[width=0.33\textwidth,angle=-90]{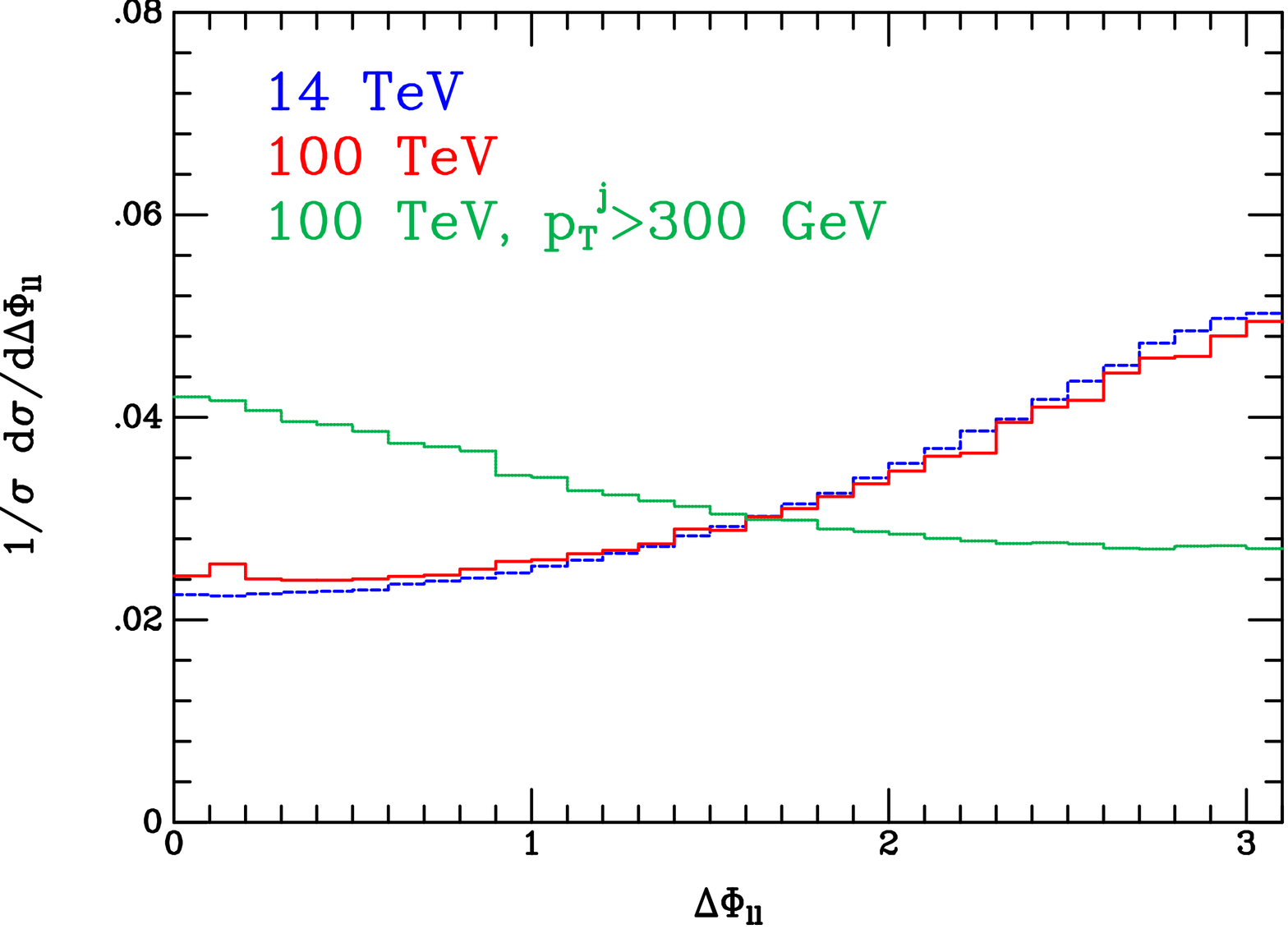} \hspace*{0.5cm}
\includegraphics[width=0.33\textwidth,angle=-90]{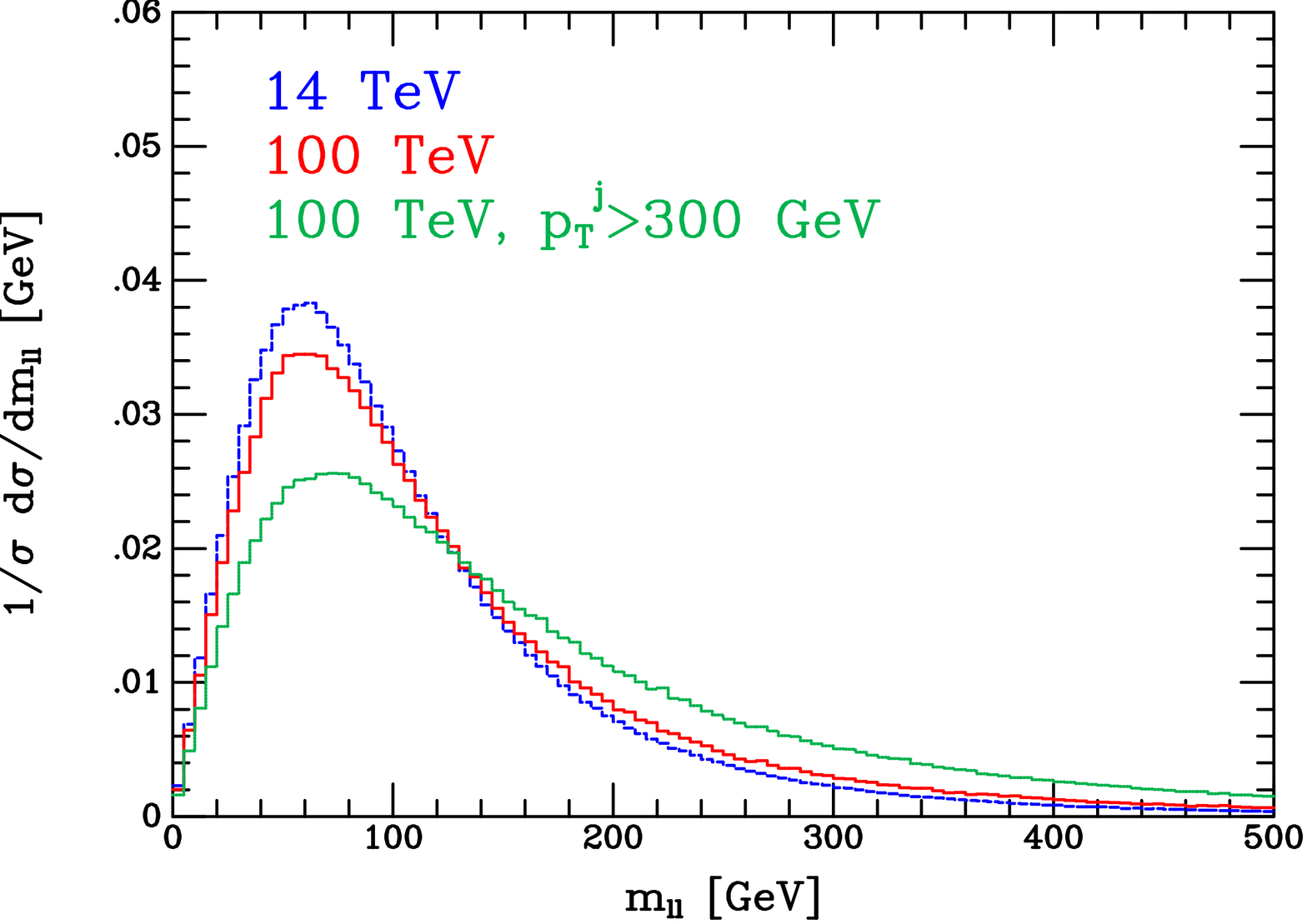}
\end{center}
\caption{NLO $\Delta \Phi_{\ell \ell}$ (left) and $m_{\ell \ell}$ (right)
distributions, normalized by the respective total cross sections, for 14 \TeV (red),
100 \TeV (blue), and 100 \TeV * (green) 
\label{fig:comp3}}
\end{figure}

Turning to leptonic observables, Fig.~\ref{fig:comp2} shows the transverse momentum and
rapidity of the positron from the $W^+$ decay.  The transverse momentum spectrum of the positron
falls much less steeply at $100$~TeV, and even less so with a higher jet cut.
The rapidity distribution of the positron is also
changed non-trivially, with the broader peak at $100$~TeV reflecting the fact that the
process is probing a much smaller parton fraction.  When the jet cut is raised to $300$~GeV
the required parton fraction is again larger so that the shape is a little closer
to the one found at $14$~TeV.~\footnote{Although not shown
here, the jet rapidity exhibits a similar behaviour.}
An observable that is particularly interesting for this process is the azimuthal angle
between the electron and the positron, which can be used to isolate contributions to this
final state from Higgs boson decays.  As shown in Fig.~\ref{fig:comp3}, under the usual
jet cuts at $14$~TeV, this distribution is peaked towards $\Delta \Phi_{\ell \ell}=\pi$,
a feature which persists at $100$~TeV using the same jet cut.  Once the jet cut is raised
significantly, the recoil of the $W^+W^-$ system results in the two leptons instead being
preferentially produced closer together, i.e. in the region $\Delta \Phi_{\ell \ell} \to 0$.
This is the same region of $\Delta \Phi_{\ell \ell}$ that is favoured by events produced
via the Higgs boson decay.
Even if the jet threshold at a $100$~TeV collider were not as high as $300$~GeV, such a shift
in this distribution could be an important consideration in optimizing Higgs-related analyses
in the $W^+W^-$ decay channel.
Despite this shift to smaller $\Delta \Phi_{\ell \ell}$, the combination of this effect
with the change in the $p_{\perp, \ell}$ distribution shown earlier results in a relatively
similar distribution for $m_{\ell \ell}$, albeit with a longer tail.

\section{Summary}
In this contribution we have considered the process $W^+\,W^-$ + jet at NLO QCD, making
use of an analytic calculation implemented into the Monte Carlo event generator
MCFM. We have considered total cross sections as well as several differential
distributions at proton-proton colliders with 14 \TeV~and 100
\TeV~center-of-mass energies.  For the latter case we have also
considered the effect of increasing the minimum $p_{\perp,j}$ cut by roughly an
order of magnitude. We found that in general at 100 \TeV~dimensionful variables
such as $p_\perp$ or $m_{\ell\ell}$ exhibit longer tails in the distributions,
reflecting the increased center-of-mass energy of the system; the increase of
the center-of-mass energy also leads to broader rapidity distributions.
Furthermore, applying a higher $p_\perp$ cut significantly changes distributions
for the dilepton azimuthal angle $\Delta \Phi_{\ell\ell}$ as well as the total
transverse momentum of the visible system $H_T$, which are frequently used for
background suppression for Higgs measurements or BSM searches, respectively. In
case such an increased cut is applied, this needs to be taken into account when
devising the respective search strategies at a 100 \TeV~machine.

%\bibliography{yrep}
%\end{document}